# Hydropathic evolution of hemagglutinin and neuraminidase glycoproteins of A(H1N1 and H3N2) viruses

J. C. Phillips

Dept. of Physics and Astronomy, Rutgers University, Piscataway, N. J., 08854

Abstract

More virulent strains of influenza virus subtypes H1N1 appeared in 2007 and H3N2 in 2011. The amino acid differences from prior less virulent strains appear to be small when tabulated through sequence alignments and counting site identities and similarities. Here we show how analyzing fractal hydropathic forces responsible for globular compaction and modularity quantifies the mutational origins of increased virulence, and also analyzes receptor sites and N-linked glycan accretion.

The two most common influenza A subtypes are H1N1 and H3N2. These viruses are mutation prolific, and their HA (hemagglutinin) and NA (neuraminidase) glycoproteins have evolved in response to migration and vaccination pressures in H1N1 [1,2]. Hydropathic analysis is based on the thermodynamics of the differential geometry of water films that encapsulate globular proteins quite generally. The availability of thousands of mutated amino acid sequences of H1N1 made it an especially attractive subject for fundamental analysis.

The emergence of a new H3N2 subtype in 1968 immediately raised the question of how one should characterize the differences between H1N1 and H3N2. Clinical studies in 1985 showed that H3N2 was more severe in all ages and had a higher infection rate in young children than H1N1 [3]. At the molecular level these differences appear to be small when tabulated through sequence alignments and counting site identities and similarities [4]. The appearance of more virulent strains of H1N1 in 2007 and of H3N2 in 2011 raised the even more difficult question of finding HA and NA sequence differences that could be responsible for enhanced virulence. Lurking in the background is the mystery of the extremely virulent 1918 H1N1 pandemic strain, with its peculiar W age mortality profile, different from the normal U profile [5]. This can be explained if a similar but hitherto unknown strain existed around 1890, or perhaps by cold and crowded conditions associated with World War I trench warfare [1].

The properties of HA are most probably dominated by binding of their glycosidic arrays and/or the conformations associated with its sialic acid receptor site. These are distributed over many amino acids, and their associated collective changes cannot be quantified only by counting site identities and similarities. There is a growing number of exemplary studies of HA glycosidation and receptor sites [6,7]. We use [7] to discuss the evolution of HA and NA and their differences.

**Methods.** In the absence of a general theory one is forced to study individual site mutations in isolation. Such comparisons are unavoidably unsatisfactory, even in the phylogenetic context (too many mutational combinations, Levinthal's paradox [8]). Evolution tends to make all proteins nearly perfect with respect to their evolving environments, and many studies have shown that hydropathic forces dominate all others in both collapsing protein chains and modularly optimizing reversible conformational changes in their globular shapes. Thermodynamically this means that proteins are near critical points, where modular amino acid effects are described by fractals. These fractals (which appear to be universal) were discovered in MZ's survey [9] of the differential surface geometry of 5226 protein segments. Hydropathicity values $\psi(n)$ are tabulated and compared to the standard KD water-air unfolding hydropathicity scale [10] in [11].

The central problems facing hydropathic scaling theory are (I) whether it is possible to describe modular conformational shape changes with $\psi(n,W)$, which is $\psi(n)$ averaged over a single sliding window length W, and (II) determining that W. (I) cannot hold exactly, but the evolutionary results for H1N1 in [1,2] were most encouraging, as large differences between HA and NA evolution of $\psi(n,W)$ were obvious. Here the evolution of H3N2 is analyzed using the prior [1,2] H1N1 unadjusted values W = 17 for NA and W = 111 for HA. Both of these H1N1 values of W are large compared to W = 1 (individual site mutations, which do not include modular effects). The superiority of large W values has been demonstrated by the evolution of lysozyme *c* [11, 12], and further examples are shown below. The H3 1968-2011 sequences used here are the same as in Table I of [7], with the addition of A/Texas/50/2012. The N2 sequences are from the Netherlands, Memphis and Hong Kong (1968,72,75,80,85,89,95,99,2002,06) and then are the same as the H3 sequences, 2007-2012.

**Results: NA roughness profiles** The benchmark result for N1 is the panoramic presentation in Fig. 6 of [1] of the opposing effects of migration and vaccination on NA roughness (variance of $\psi$(aa,17), MZ scale) averaged over many strains. The data base for N2 is smaller, but similar trends are seen in Figs. 1(a,b) here. A surprising feature is less scatter using the KD scale, which appears to be more reliable for N2. The more virulent strain, first seen in Victoria 2011, reverses the KD reduction in vaccination-induced roughness from 1968 up to 2009, and this increase has continued from 2012 to 2014. It is consistent with the increased virulence of the new strains, but the increase in KD17 roughness so far is modest compared to the reduction since 1968.

**HA** is composed of two chains, HA1 (18-342) and HA2 (344-565) and a signal peptide (1-17). Here we focus on 1-343, and our site numbering differs from normal numbering of HA1 by a shift of 17. A simple, primarily monotonic evolutionary pattern was not found for H1 in [2]. Instead reassortment features (‘proteinquakes”) of the sialic acid receptor site 130-230 on the HA1 head were studied in detail, especially in connection with the appearance of a new virulent strain, first in swine around 2000 and then developing in humans in successive large cities up to 2007. This strain was rendered less virulent after 2009 by a widespread vaccination program. This H1 discussion can be greatly improved for H3, both for the receptor site (Fig. 3a, and for PNG’s N-linked glycans, Fig. 4a of [7]).

**N-linked glycans** There are four conserved PNG’s (39, 55, 183, 303) in HA 1-343, and seven have been added since 1968 (80, 143, 263, {139, 150, 161}, 62). The glycan enrichment of the HA1 head of H3 is shown in Fig. 1b of [7]. The closely spaced triplet {139, 150, 161} appeared abruptly between 1995 and 1998. Fig. 2(a,b) shows the W= 111 hydroprofiles of <$\psi$(j)W> for two strains from Memphis, 1995 and 1999, with a large hydrophilic shift between sites 100 and 210. The conserved site 183 is located in the deepest hydrophilic 1995 minimum in Fig. 2(a). The 1995 and 1999 regions between 120 and 200 are shown enlarged in Fig. 2(b). The triplet sites of {139, 150, 161} are located at minima, which have deepened from 1995 to 1999, so that at these new j sites <$\psi$(j)W> is lower in 1999 than <$\psi$(183)W> was in 1995.

This striking quadruplet correlation with hydrophilic minima is unlikely to be accidental. It holds in the receptor binding domain, and a possible explanation is that in the thicker (more hydrophilic protein) regions the water film acts as a soft spacer between sialic acid or the antibody fragment F045-092 and PNG's, see for example Fig. 4(b) of [7]. The results are best with the MZ scale [9] and W ~ 100 or more. The KD scale [10] (correlation 0.85 with the MZ scale [11]) fails, as it makes 150 (not 180) the deepest 1995 hydrophilic minimum (not shown here). This is a striking example of the superiority of the MZ scale, which appears to be universal for proteins larger than 100 aa [12,13].

**Receptor Site** [7] probed the receptor site with the antibody fragment F045-092, which has the smallest epitopic footprint to date. In some ways the antibody is an excellent site probe, perhaps better than sialic acid itself, for the latter interacts both with proximal sites (by contacts) and with distal sites (through protein strain). The footprint of F045-092 is more than twice that of sialic acid and may include not only conserved proximal sites, but also most of the variable distal sites. This is consistent with the success of F045-092 in binding to a wide range of H3 strains, from 1963 to 2011 [7].

The F045-092 conserved contact residues on HA, which are shown in Fig. 3(a) of [7], consist mainly of three segments, A: 148-154, B: 170-175, and C: 210-211. Two strains were studied [7], Victoria 1975 and 2011, with substantial reassortment of chain HA1 (82% Identities, 90% Positives). The hydropathic profile changes of these two strains in Fig. 3(a) show a simple block pattern, with hydrophobic stabilization near the ends and center blocks, and hydro-philic, -phobic inversion in the border blocks 80-150 and 175-250. Specifically the stable narrow central block contains the A and B segments of the receptor site, which is little changed compared to the border regions. This is consistent with the observed universal binding of F045-092 to H3 subtypes [7].

The differences between the new virulent 2011 H3 strain and the old 1975 strain near the central receptor site are shown in Fig. 3(b). The receptor site is stabilized by the two hydrophobic internal pinning peaks, while the surface segments A and B lie on the outsides of these two internal peaks, and connect to the outside hydrophilic minima. Evolution from 1975 to 2011 has a striking effect: it levels not only the A and B peaks, but also the C peak, as shown by the double arrows. This leveling effect [14] is important to understanding protein dynamics in the context of amino acid surface areas. It has been identified in cyclooxygenase, where it distinguishes between two isozymes encoded by separate genes [15]. The deep hydrophilic valley between 170 and 190 facilitates the fold that enables C to synchronize dynamically with A and B in the receptor site. These favorable features (leveling and hinging) are absent from the profiles calculated with the 1982 KD scale. This is not surprising, as the KD scale is based simply on experimentally measured water-air unfolding enthalpies, and not on surface areas derived by Voronoi partitioning.

**More virulent 2011-2012 strain** How do we explain the sudden emergence of this more virulent H3N2 strain? It is probably not caused by the accretion of the last glycan, as Fig. 4(a) of [7] shows that this occurred gradually, starting in 2008. The shift of the border region 175-250 did not occur gradually between 1975 and 2011, as one might suppose from Fig. 3(a). As Fig. 4 shows, the shift occurred abruptly between Perth 2009 and Victoria 2011, two strains with 98% sequence identity. It can be traced chiefly to two closely spaced Ser mutations, S230I and S235Y, labeled D in Figure 4.

On the MZ scale [11] Ser is the most hydrophilic amino acid above Glu, Asp, Arg and Lys, whereas Ile and Tyr are just below Cys and Val as the most hydrophobic amino acids. The 175-250 block shift aligns the hydrophobic peaks associated with segment C: 210-211 and D: 230,235 with the corresponding A: 148-154 and B: 170-175 pinning peaks. As a check on this mechanism, there are many reported 2012-2014 sequences containing these two mutations, indicative of more virulence. Singapore/GP1063/2011 contains only one (S230I, but not S235Y), and this combination appears in no other reported H3 sequence.

There is a surprise hidden in the leveling of A-D. The interaction range associated with W = 111 is 55, but A is ~ 80 sites away from D. Physical scientists might be tempted to describe this interaction as "indirect", and mediated by a chain of A-B, B-C, C-D interactions. However, given the critical nature of core interactions, it is probably better to say that the four segments belong to a common dynamical set in the most virulent 2011-2012 strain. This case shows that mutations of even distal amino acids can generate delicate extremum leveling effects that alter viral dynamics by creating new sets. These effects are caused by the folding of hydrophobic segments inside the protein core, which affects the balance of hydrophilic surface segments that determine the shape and dynamics of the receptor site. Similar thermodynamic balancing and synchronizing mechanisms have been invoked to explain the dynamics of motile cells [16], and they guide modular interactions in evolving protein networks [17].

**Relation to H1N1** The more common influenza strains of H1N1 have undergone many mutations, discussed for N1 in [1] and H1 in [2]. In the light of the present results, as well as the similarities in HA receptor sites seen by comparing Fig 4A of [18] with Fig. 3(a) of [7], one can ask whether any of the H1 strains exhibit similar leveling of the A,B,C hydrophobic peaks, as in Figs. 3(b) and 4 here. Looking at Fig. 6 of [2], we see just such a strain, labeled 1950 Fort Warren, when the US Army vaccination program abruptly became more effective. Overall, the deep hydrophilic valley found in the center of H1 HA1 is much more shallow in H3, suggesting increased virulence of H3.

**A broader picture** The main emphasis here has been on the leveling effects shown in Figs. 3(b) and 4, as these have appeared abruptly and mysteriously in H3N2 in 2011. Such effects are exceptional. The broader historical trends [3] for H1 and H3 are shown in Figs. 5 and 6 and discussed in their figure captions. The largest difference between H1 and H3 is the ordinate ranges of $\psi(n,111)$ in the figures – 139-154 for H1, and 145-154 for H3, meaning overall H3 is more hydrophobic. Increased hydrophobicity tends to stabilize the glycoprotein and facilitate oligomer formation, increasing virulence.

**Discussion** While the general problems of protein folding, protein dynamics, and protein-protein binding remain unsolved [8,19,20], with a sufficiently strong data base (such as [7] provides) one can identify factors involved in clinically important viral reassortments, an extremely complex network problem. Analytical tools of the kind developed here have helped to guide the semiconductor opto-electronic revolution (computers and the internet), and interdisciplinary methods could be useful in engineering protein networks as well. They have already helped change the technology of glass networks from art to science [21,22]. Until this post-2005 aluminosilicate work of Corning, glass technology had shown few changes since silicates were extended to include borosilicates in the late 19th century.

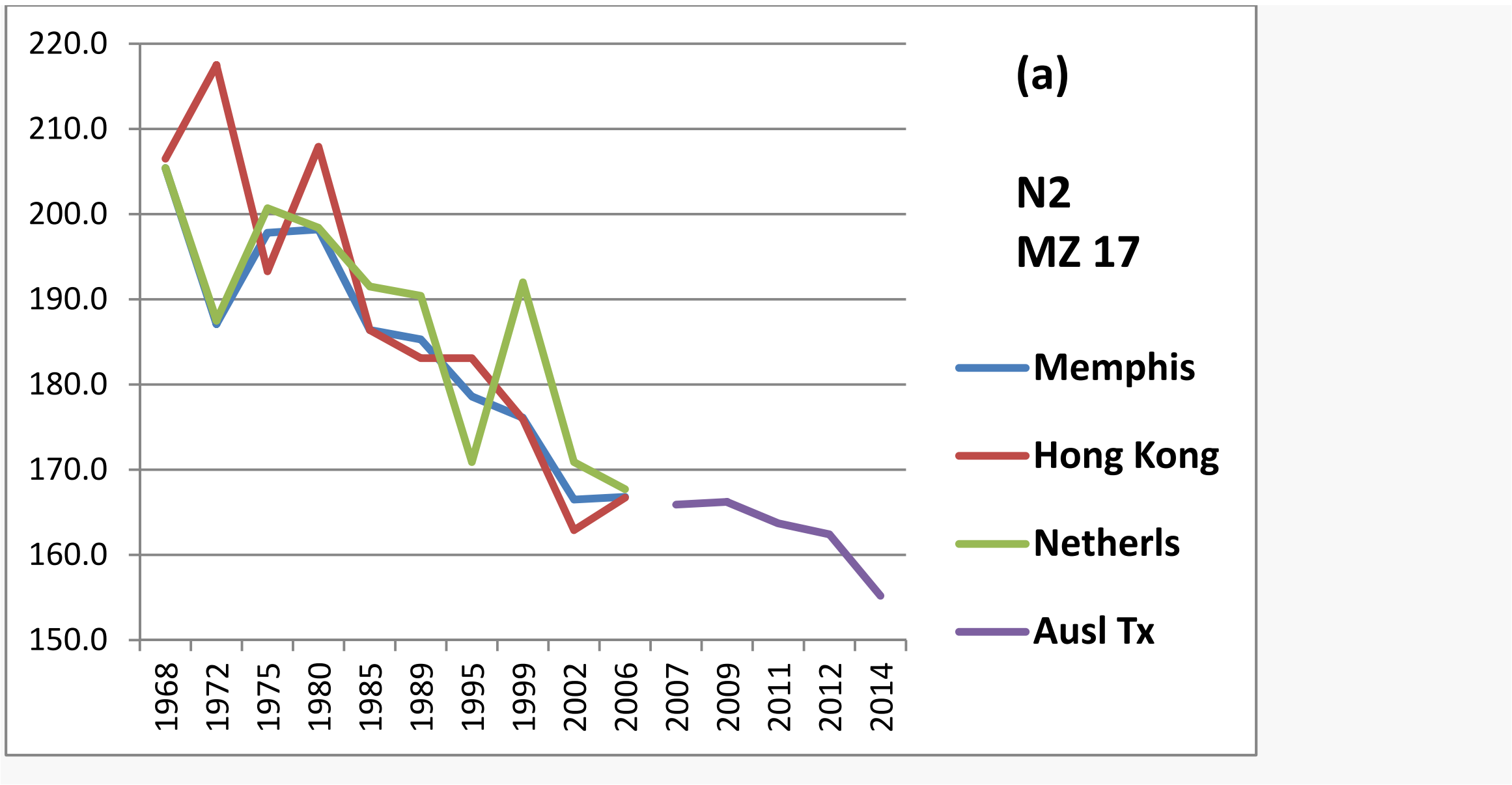

Fig. 1(a) Evolution of H3 roughness (variance of ψ(n, W) with W = 17 for NA, using the MZ scale. The main feature here is the broad downward trend, which is not obscured by regional fluctuations.

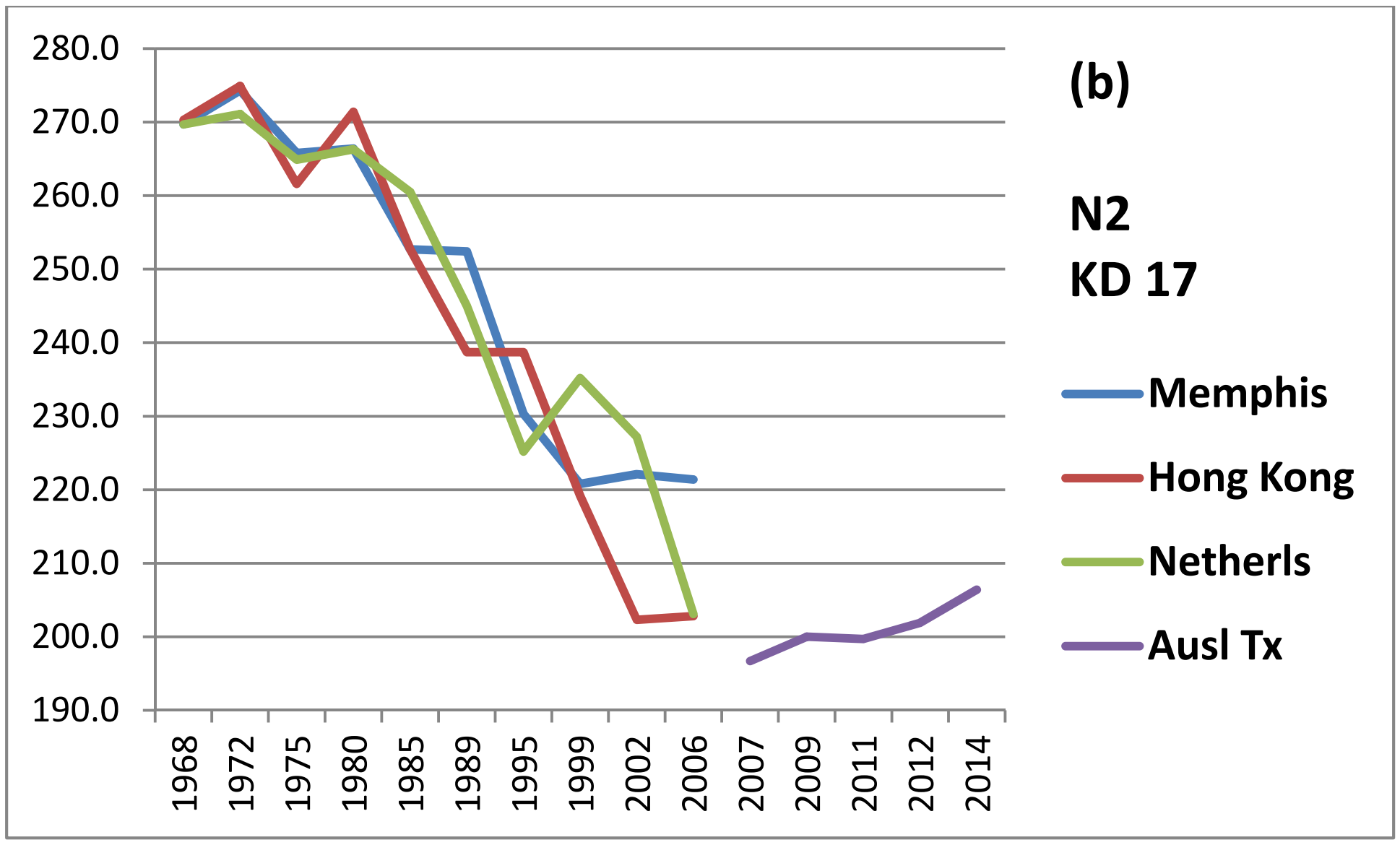

Fig. 1(b) Evolution of H3 roughness (variance of ψ(n, W) with W = 17 for NA, using the MZ scale. Here the regional fluctuations are smaller. Combining .this figure with 1(a) suggests that the downward trend may have leveled off with the appearance of a virulent new strain in 2011-2012.

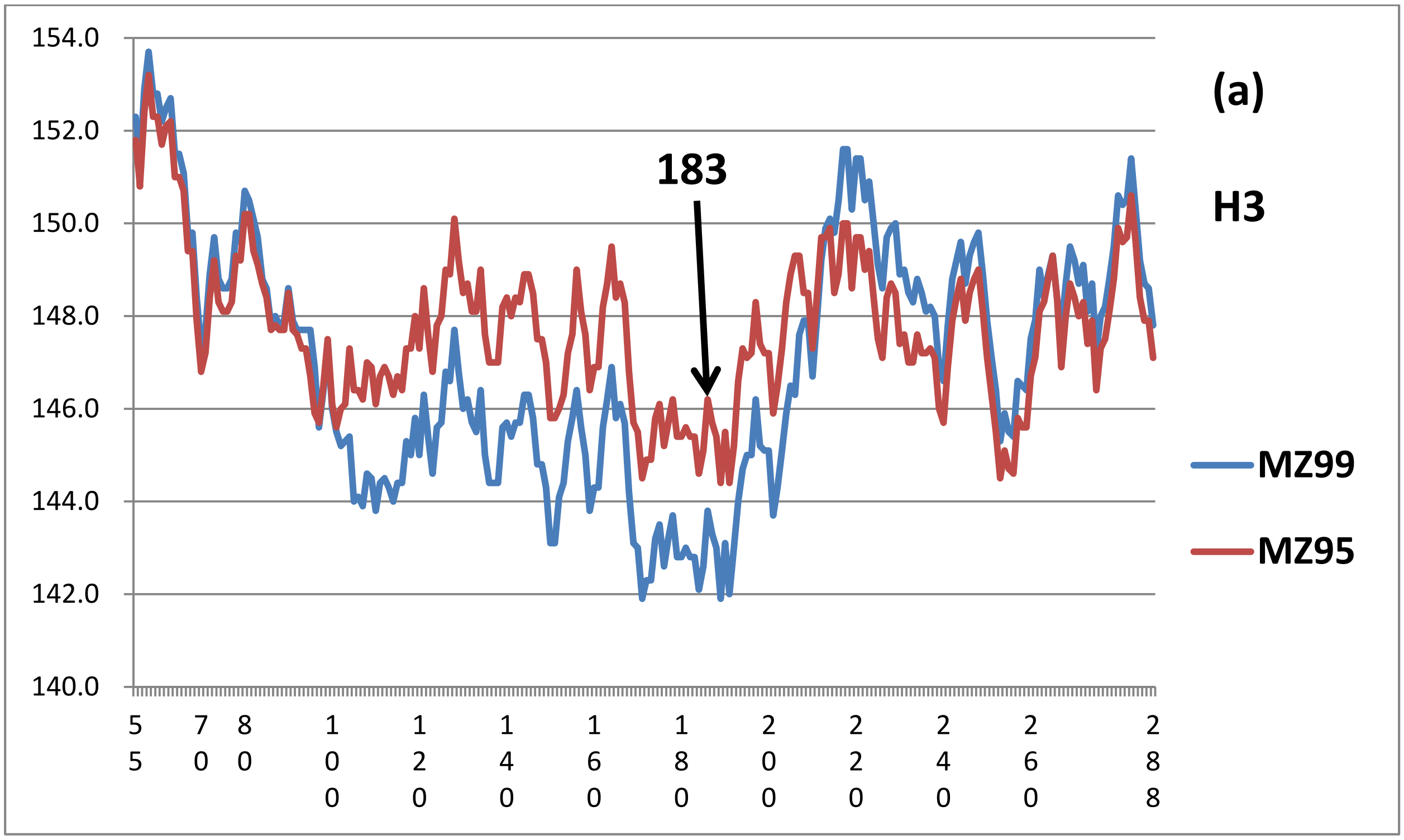

Fig. 2(a).The conserved glycosite 183, already present in 1968, lies close to the receptor site and is near the center of the HA1 chain 17-343.

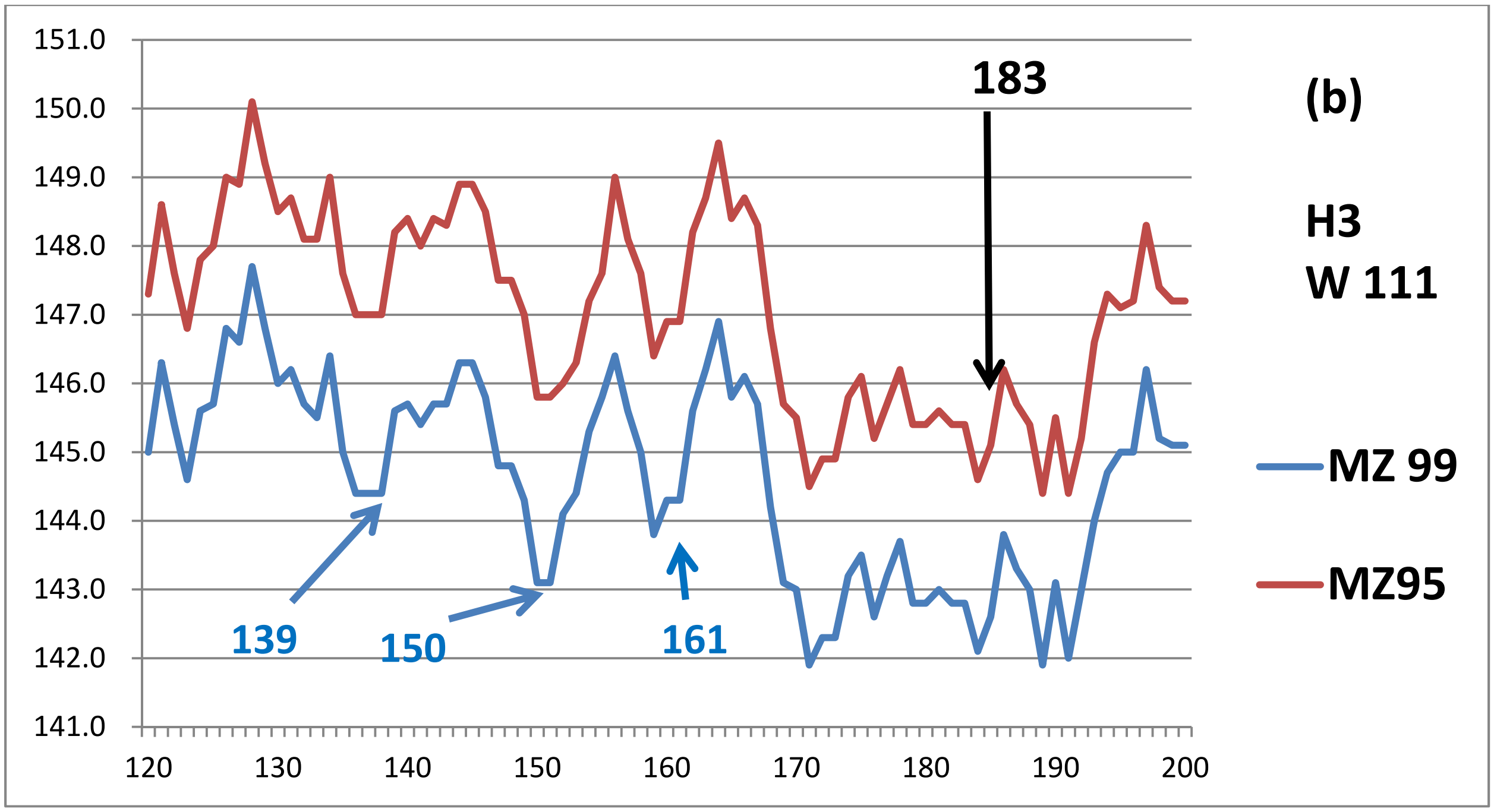

Fig. 2(b) The burst of three new glycosites (see Fig. 4(a) of [6]) between 1995 and 1999, is localized near successive hydrophilic minima of ψ(n,W = 111) of H3. Using the MZ scale, in 1999 these minima drop below the value of ψ(183,111) in 1995.

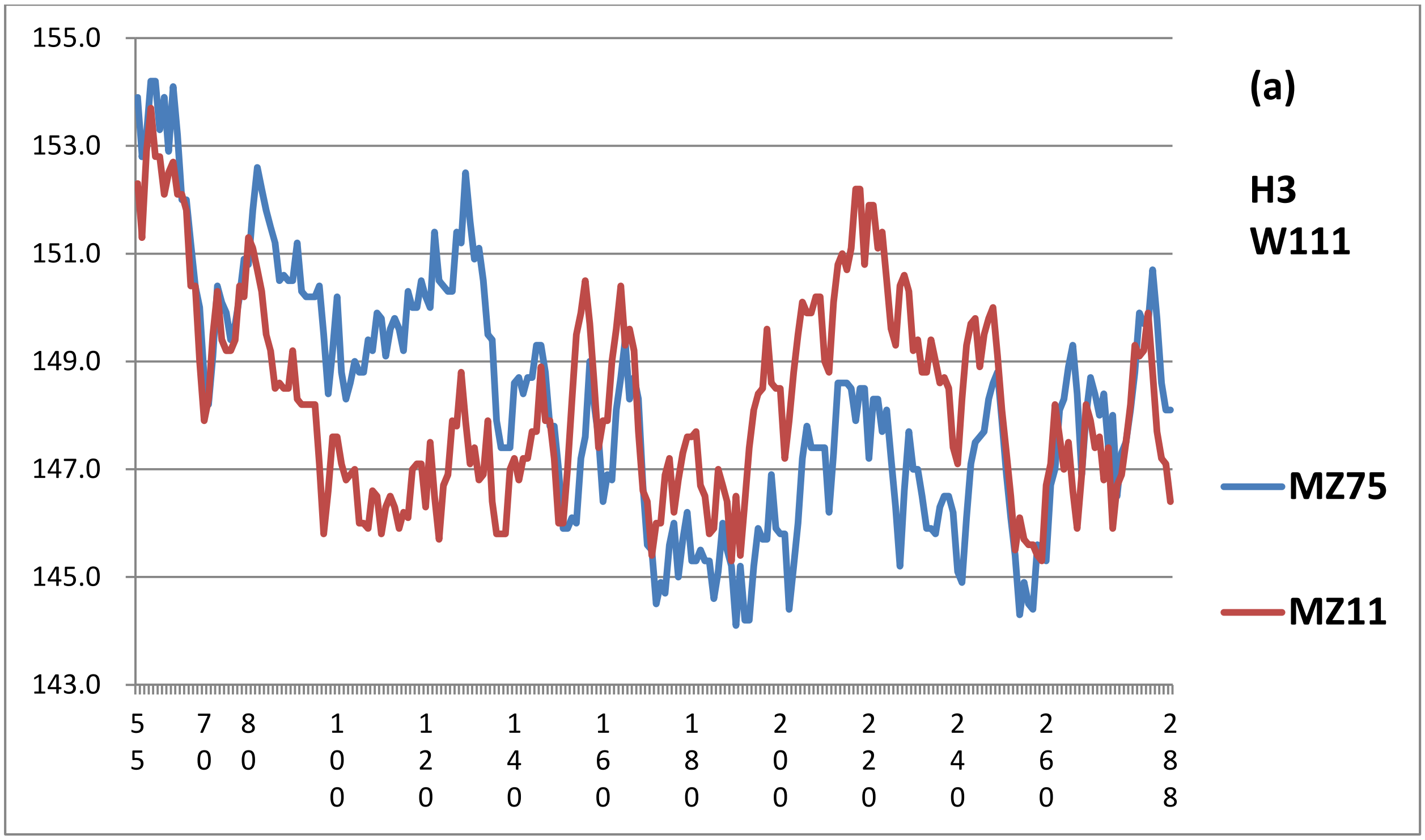

Fig. 3(a). The overall panoramas of ψ(n,111) for Victoria 1975 and 2011, with the MZ scale.

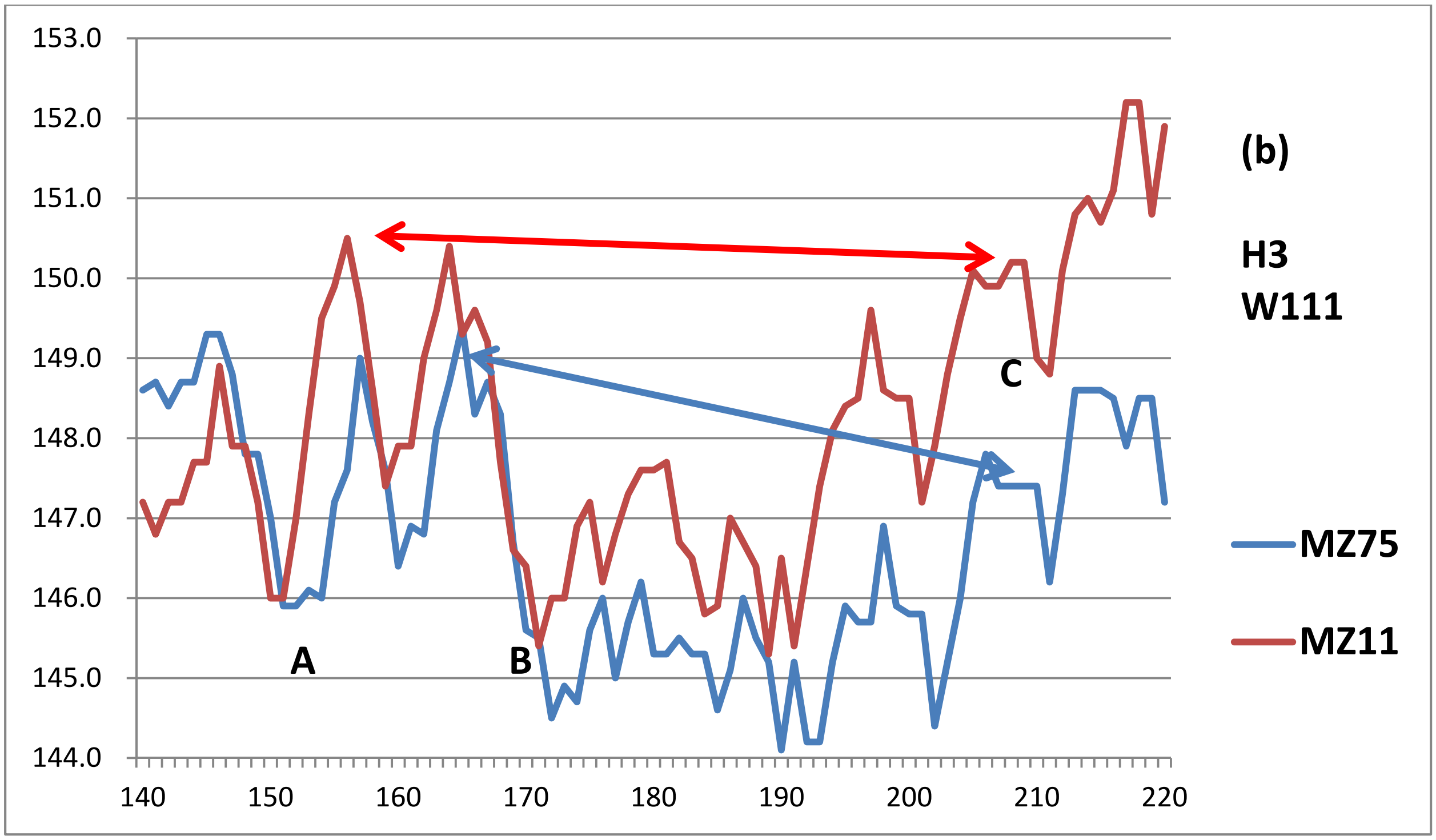

Fig. 3(b) The receptor region containing the A, B, C segments is enlarged from the overall Fig. 3(a). As explained in the text, the increased virulence of the new H3N2 stain is attributed to the leveling of the A,B,C hydrophobic pinning peaks (shown by the double arrows) in the new strain.

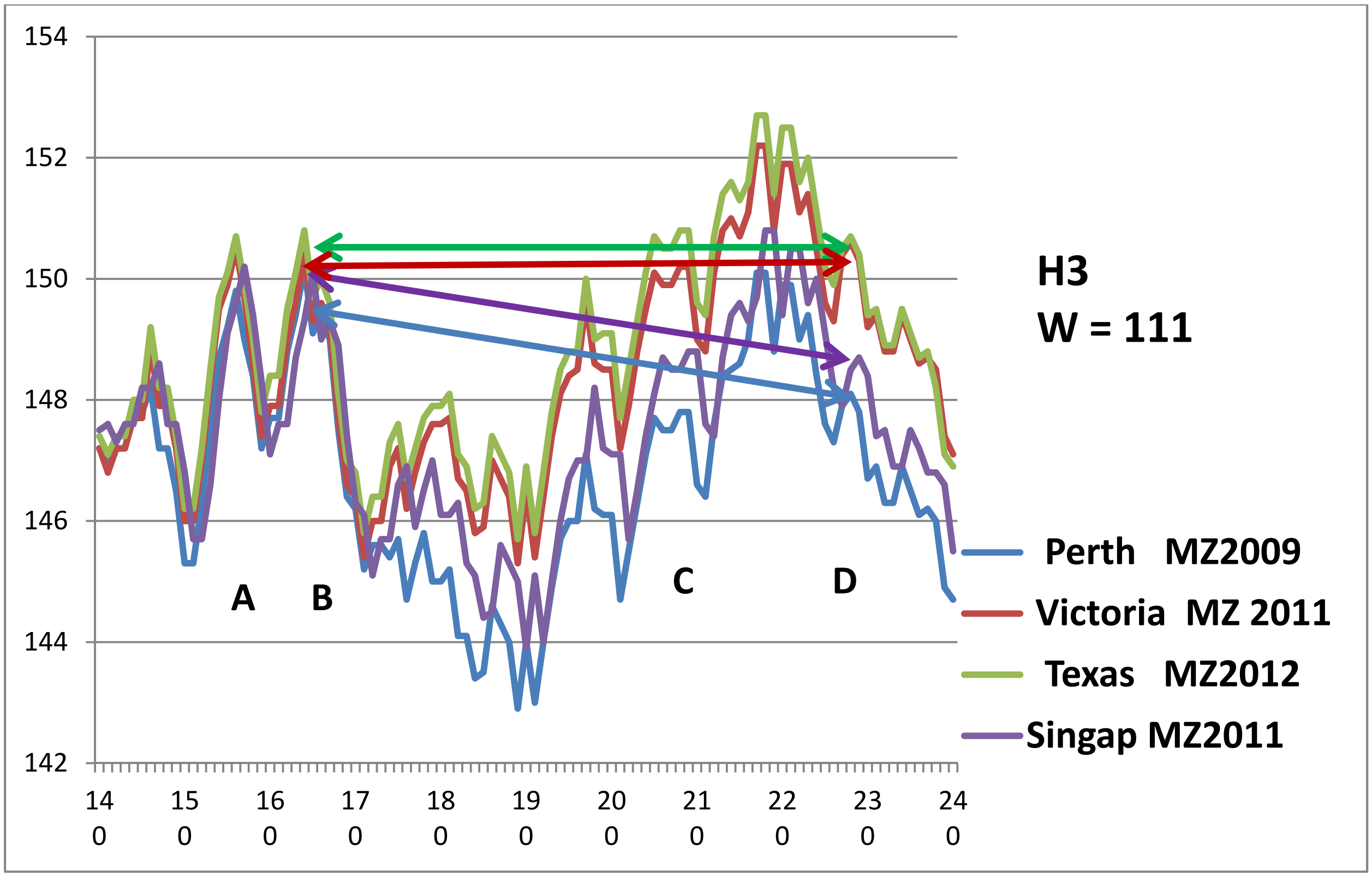

Fig. 4. Hydroprofiles of ψ(n,W) for H3 in the n range including the A,B,C receptor segments, and the D mutated pair. The A-D hydrophobic pinning points are level for Victoria 2011 and Texas 2012, which contain the two key D mutations S230I and S235Y, but not for Singapore 2011 (missing S235Y) and Perth 2009 (both missing).

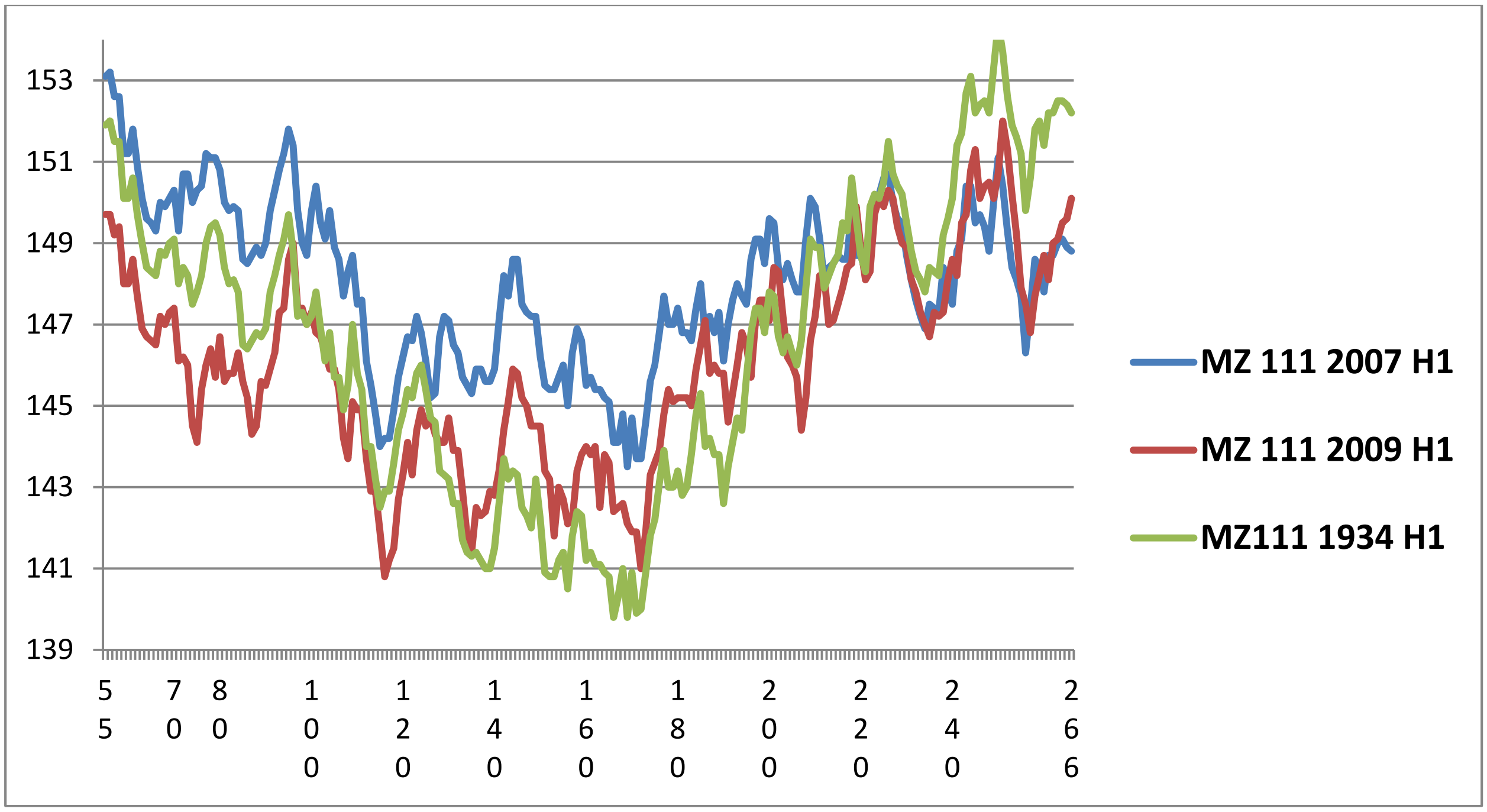

Fig. 5. Broad historical trends in the evolution of the HA1 chain of H1 from H1N1 (details in [2]). The earliest unreconstructed strain, Puerto Rico 1934, shows a simple "V" structure, with a hydrophilic hinge near the center of the 130-230 sialic acid binding site. The swine flu strain (Texas, 2007) is much more hydrophobic in the receptor site, and also towards the N terminal, presumably enhancing its virulence. By 2009 a widespread vaccination program had reinstated milder strains circulating before 2000, which are more hydrophilic than Puerto Rico 1934 [1,2].

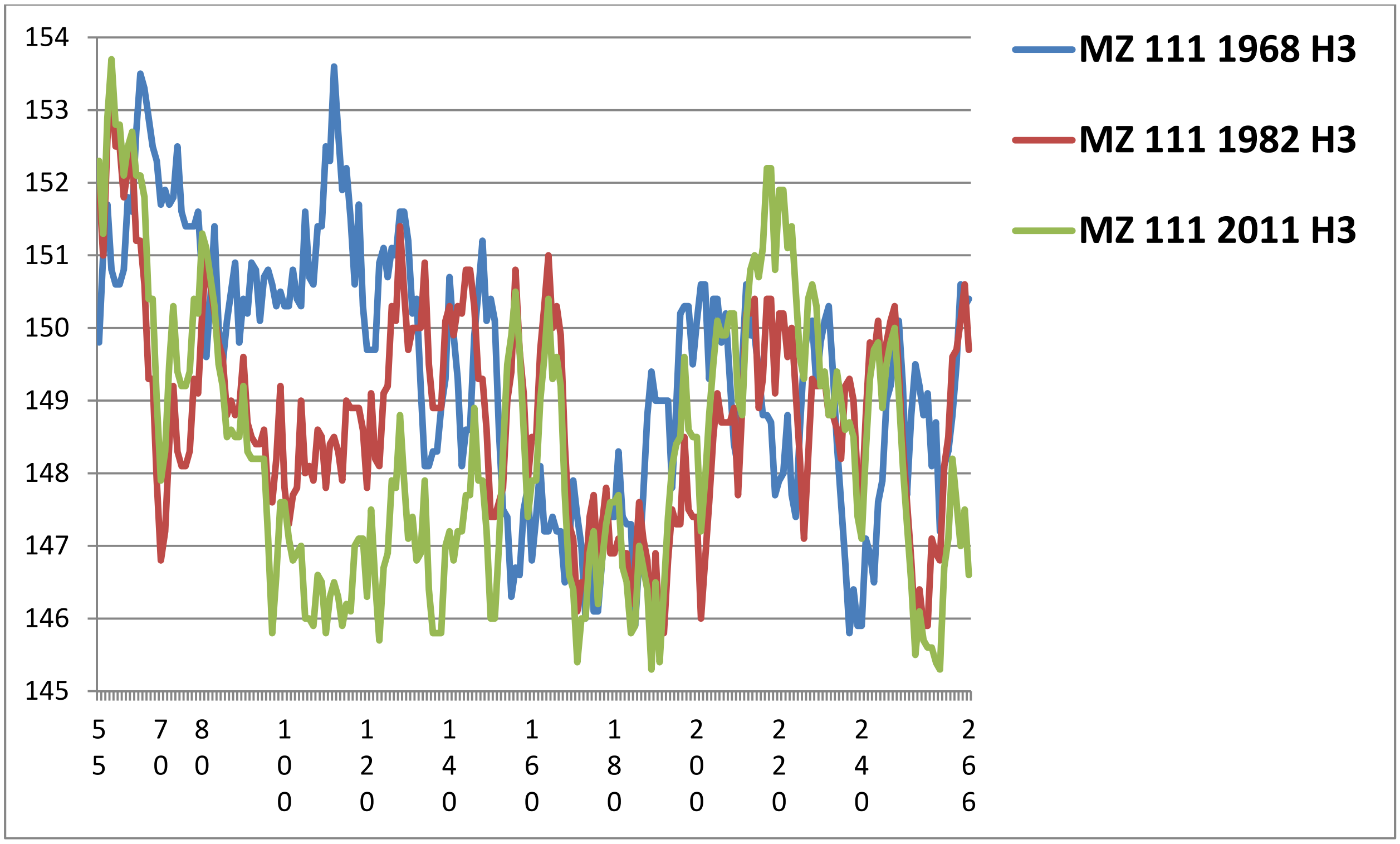

Fig. 6. At its first appearance in Hong Kong in 1968, H3 had a more complex and more virulent profile than H1 2007, because of larger hydrophobicity and stabilization of the receptor site by a strong hydrophobic peak near 115. In the 1982 Hong Kong strain this peak had disappeared, perhaps because it made strains less infectious. The trend towards increased hydrophilicity near 115 continued in Victoria 2011, but a new hydrophobic peak appeared on the C-side of the receptor site near 230. This new peak is associated with the leveling effects discussed in Figs. 3(b) and 4, which made H3 more virulent again.